\documentclass[sigconf]{acmart}
\usepackage[T1]{fontenc}
\usepackage{aecompl}
\usepackage{amsmath,amsfonts}
\usepackage{algorithmic}
\usepackage{graphicx}
\usepackage{xcolor}
\usepackage{hyperref}
\usepackage{setspace}
\usepackage{booktabs} 
\usepackage{subfigure}
\usepackage{amsmath}
\usepackage{color}
\usepackage{amsmath}
\usepackage{mathrsfs}
\usepackage[normalem]{ulem}
\usepackage{enumitem}
\usepackage{multirow}
\usepackage{ulem}
\usepackage[nomargin,inline,marginclue,draft]{fixme}
\usepackage{balance}
\usepackage{verbatim}
\usepackage{diagbox}
\usepackage{changepage}
\usepackage{xcolor}
\usepackage{bm}
\usepackage{makecell}

\newsavebox\CBox
\def\textBF#1{\sbox\CBox{#1}\resizebox{\wd\CBox}{\ht\CBox}{\textbf{#1}}}

\DeclareMathOperator*{\concat}{\scalebox{1}[1.5]{$\parallel$}}

\definecolor{myred}{rgb}{1, 0, 0}
\definecolor{myblue}{rgb}{0, 0, 1}
\definecolor{myblack}{rgb}{1, 1, 1}

\newcommand{\bs}[1]{\boldsymbol{#1}}

\newcommand{\para}[1]{{\vspace{4pt} \bf \noindent #1 \hspace{0pt}}}

\newlength\savedwidth

\copyrightyear{2021}
\acmYear{2021}
\setcopyright{acmcopyright}\acmConference[SIGIR '21]{Proceedings of the 44th International ACM SIGIR Conference on Research and Development in Information Retrieval}{July 11--15, 2021}{Virtual Event, Canada}
\acmBooktitle{Proceedings of the 44th International ACM SIGIR Conference on Research and Development in Information Retrieval (SIGIR '21), July 11--15, 2021, Virtual Event, Canada}
\acmPrice{15.00}
\acmDOI{10.1145/3404835.3462968}
\acmISBN{978-1-4503-8037-9/21/07}

\begin{document}

\title{Sequential Recommendation with Graph Neural Networks}
\author{Jianxin Chang$^{1}$, Chen Gao$^{1}$, Yu Zheng$^{1}$, Yiqun Hui$^{2}$, Yanan Niu$^{2}$, Yang Song$^{2}$, Depeng Jin$^{1}$, Yong Li$^{1}$}
\affiliation{\institution{$^{1}$Beijing National Research Center for Information Science and Technology (BNRist),\\
Department of Electronic Engineering, Tsinghua University}
  }
\affiliation{\institution{$^{2}$Beijing Kuaishou Technology Co., Ltd.}}
\email{liyong07@tsinghua.edu.cn}
\renewcommand{\shortauthors}{Chang~\textit{et al.}}

\begin{abstract}
Sequential recommendation aims to leverage users' historical behaviors to predict their next interaction.
Existing works have not yet addressed two main challenges in sequential recommendation. 
First, user behaviors in their rich historical sequences are often implicit and noisy preference signals, they cannot sufficiently reflect users' actual preferences.
In addition, users' dynamic preferences often change rapidly over time, and hence it is difficult to capture user patterns in their historical sequences.
In this work, we propose a graph neural network model called SURGE 
(short for \textit{\textBF{S}eq\textBF{U}ential \textBF{R}ecommendation with \textBF{G}raph neural n\textBF{E}tworks}) 
to address these two issues.
Specifically, SURGE integrates different types of preferences in long-term user behaviors into clusters in the graph by re-constructing loose item sequences into tight item-item interest graphs based on metric learning.
This helps explicitly distinguish users' core interests, by forming dense clusters in the interest graph.
Then, we perform cluster-aware and query-aware graph convolutional propagation and graph pooling on the constructed graph. It dynamically fuses and extracts users' current activated core interests from noisy user behavior sequences.
We conduct extensive experiments on both public and proprietary industrial datasets. Experimental results demonstrate significant performance gains of our proposed method compared to state-of-the-art methods.
Further studies on sequence length confirm that our method can model long behavioral sequences effectively and efficiently.
\end{abstract}

\maketitle
\renewcommand{\thefootnote}{\fnsymbol{footnote}}

\section{Introduction}\label{sec::intro}
Sequential recommendation attempts to predict a user's next behavior by exploiting their historical behavior-sequences, 
which has been widely adopted in modern online information systems, such as news, video, advertisements, etc.
Differ from traditional recommendation tasks that model user preferences in a static fashion, sequential recommendation is capable of capturing user's evolved and dynamic preferences.
For example, a user may prefer to watch soccer news only during the period of World Cup, which can be regarded as a kind of \textit{short-term} preference.

Existing works have realized the significance of modeling fast-changing short-term preferences, by approaching the problem from three perspectives.
Specifically, early efforts~\cite{rendle2010factorizing, DIN} adopt human-designed rules or attention mechanism to assign time-decaying weights to historically interacted items. 
The second category of works~\cite{DIEN, GRU4REC} leverages recurrent neural networks to summarize the behavioral sequences, 
but they suffer from the short-term bottleneck in capturing users' dynamic interests
due to the difficulty of modeling long-range dependencies.
Recent solutions~\cite{SLIREC} jointly model long-term and short-term interests to avoid forgetting long-term interests, but the division and integration of long/short-term interests are still challenging.
In short, the aforementioned works commonly concentrate more on user behaviors of recent times, and are not capable of fully mining older behavior-sequences to accurately estimate their current interests.
As a result, there are two major challenges in sequential recommendation that have not been well-addressed so far as follows.

\vspace{-0.1cm}
\begin{itemize}[leftmargin=*]
	\item \textbf{User behaviors in long sequences reflect implicit and noisy preference signals.} 
    Users may interact with many items with implicit feedback, such as clicks and watches. 
	Unlike explicit feedback that can infer user preferences, such as likes and favorites, single implicit feedback cannot reflect user preferences. The user may click on items that are not of their interest most of the time and will not choose similar items for interaction afterward. However, these records will serve as noises in the user's behavior history, worsening the modeling of their real interests.
	\item \textbf{User preferences are always drifting over time due to their diversity.} 
    As we have mentioned, user preferences are changing, no matter slow or fast. Given a point of time, some preferences may be still activated and some others may have been deactivated. Thus, even if we have extracted user preferences from the implicit and noisy behaviors, it is still challenging to model how they change in the history and estimate the activated preferences at the current time, which is the core of recommendation models.
\end{itemize}

To address these two challenges, we propose a graph-based method with graph convolutional networks to extract implicit preference signals. The dynamic graph pooling is then used to capture the dynamics of preferences.
Specifically, we first convert the loose item sequence to a tight item-item graph and design a attentive graph convolutional network that gather weak signals to strong ones that can accurately reflect user preferences. We then propose a dynamic graph pooling technique that adaptively reserves activated core preferences for predicting the user's next behavior.

To summarize, the contributions of this paper are as follows,
\begin{itemize}[leftmargin=*,partopsep=0pt,topsep=0pt]
\setlength{\itemsep}{0pt}
\setlength{\parsep}{0pt}
\setlength{\parskip}{0pt}
	\item We approach sequential recommendation from a new perspective by taking into consideration the implicit-signal behaviors and fast-changing preferences.
	\item We propose to aggregate implicit signals into explicit ones from user behaviors by designing graph neural network-based models on constructed item-item interest graphs. Then we design dynamic-pooling for filtering and reserving activated core preferences for recommendation.
	\item We conduct extensive experiments on two large-scale datasets collected from real-world applications. The experimental results show significant performance improvements compared with the state-of-the-art methods of sequential recommendation. 
	Further studies also verify that our method can model long behavioral sequences effectively and efficiently.
\end{itemize}


\vspace{-3px}
\section{Problem Formulation}\label{sec::profdef}

Here we provide a formal definition of sequential recommendation.
Assume we have a set of items, denoted by $\mathcal{X}$, where $x \in \mathcal{X}$ denotes an item. The number of items is denoted as $|\mathcal{X}|$. 
Generally, a user has a sequential interaction sequence with items: $\{x_1, x_2, \ldots, x_n\}$, where $n$ is the number of interactions and 
$x_i$ is the $i$-th item that the user has interacted with. 
Sequential recommendation aims to predict the next item $x_{n+1}$ that matches the user's preferences. 
The user's preferences can be inferred from chronological user-item implicit feedback.
Based on the above definition, the task of sequential recommendation can be formulated as follows:

\noindent
\textbf{Input:} The interaction history for each user $\{x_1, x_2, \ldots, x_n\}$.

\noindent
\textbf{Output:} A recommendation model that estimates the probability 
that a user with interaction history $\{x_1, x_2, \ldots, x_n\}$ will interact with the target item $x_t$ at the $(n + 1)$-th step.

\vspace{-3px}
\section{Methodology}\label{sec::method}

\begin{figure*}[t]
\begin{center}
  \includegraphics[width=15cm]{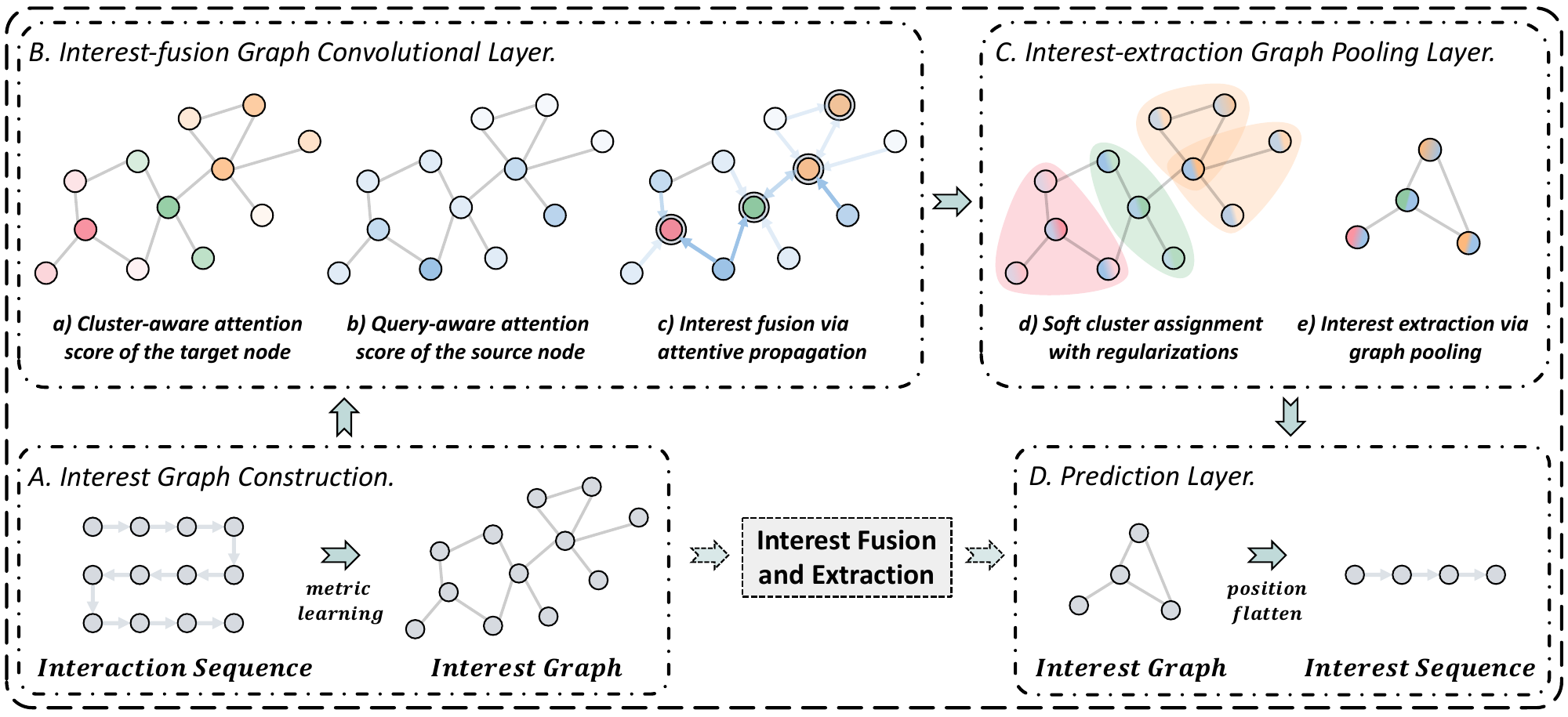}
\end{center}
\caption{Illustration of the SURGE model.
Each interaction sequence is re-constructed into an interest graph (A) based on metric learning, and interest fusion (B) and extraction (C) are dynamically performed on the graph.
The currently activated core interest sequence (D) is obtained by flattening the pooled graph after interest fusing and extracting, which can be used for further modeling and prediction.
Best viewed in color.
}\label{fig::framework}
\vspace{-0.2cm}
\end{figure*}

Figure~\ref{fig::framework} illustrates our proposed SURGE model which is made up of the following four parts, which we will elaborate on one by one.

\begin{itemize}[leftmargin=*]
  \item \textBF{Interest Graph Construction.}
    By re-constructing loose item sequences as tight item-item interest graphs based on metric learning, we explicitly integrate and distinguish different types of preferences in long-term user behaviors.

  \item \textBF{Interest-fusion Graph Convolutional Layer.}
    The graph convolution propagation on the constructed interest graph dynamically fuses the user's interests, strengthening important behaviors, and weakening noise behaviors.

  \item \textBF{Interest-extraction Graph Pooling Layer.}
    Considering users' different preferences at different moments, a dynamic graph pooling operation is conducted to adaptively reserve 
    dynamically-activated core preferences.

  \item \textBF{Prediction Layer.}
    After the pooled graphs are flattened into reduced sequences, we model the evolution of the enhanced interest signals and predict the next item that the user has high probability to interact with.

\end{itemize}

\subsection{Interest Graph Construction}
To integrate and distinguish different types of preferences in users' rich historical behaviors, we can convert loose item sequences into tight item-item interest graphs.
The co-occurrence relationship between two items is a reasonable construction criterion, but the challenge is that the sparseness of the co-occurrence relationship is not enough to generate a connected graph for each user.
In this section, we propose a novel way based on metric learning to automatically construct graph structures for each interaction sequence to explore the distribution of its interests.
  
\subsubsection{\textbf{Raw graph construction.}}

This novel module attempts to construct an undirected graph $\mathcal{G} = \{\mathcal{V}, \mathcal{E}, A \}$ for each interaction sequence, 
where $\mathcal{E}$ is the set of graph edges to learn and $A \in \mathbb{R}^{n \times n}$ denotes the corresponding adjacency matrix. 
Each vertex $v \in  \mathcal{V}$ with $|\mathcal{V}| = n$ corresponds to 
an interacted item  (and the associated embedding vector 
is denoted as
$\vec{h} \in \mathbb{R}^d$).
We aim to learn the adjacency matrix $A$, where each edge $(i,j, A_{i,j}) \in \mathcal{E}$ indicates whether item $i$ is related to item $j$.

By representing each user's interaction history as a graph, it is easier to distinguish his/her core and peripheral interests.
The core interest node has a higher degree than the peripheral interest node due to connecting more similar interests, and the higher frequency of similar interests results in a denser and larger subgraph.
In this way, a priori framework is constructed, that is, 
neighbor nodes are similar, and dense subgraphs are the core interests of users. 

\subsubsection{\textbf{Node similarity metric learning.}}
Since we need a priori graph in which neighbor nodes are similar, the graph learning problem can be transformed into node similarity metric learning, which will be jointly trained with the downstream recommendation task.
This graph construction method 
is general, easy to implement, and able to perfectly cope with inductive learning (with new items during testing). 
Metric learning can be classified into kernel-based and attention-based methods~\citep{zhu2021deep}.
Common options for kernel-based methods include cosine distance~\citep{Zhang:2020tu,Wang:2020bs}, Euclidean distance~\citep{Yu:2020vw,Zhao:2021vr} and Mahalanobis distance~\citep{Li:2018wu,Wu:2018gy}.
A good similarity metric function is supposed to be learnable to improve expressiveness and have acceptable complexity.
To balance expressiveness and complexity, we adopt weighted cosine similarity~\citep{IDGL,Zhao:2021vn} as our metric function formulated as follows,
\begin{equation}\label{eq:weighted_cosine}
\begin{aligned}
  M_{ij} = \text{cos}(\vec{\mathbf{w}} \odot \vec{h}_i, \vec{\mathbf{w}} \odot \vec{h}_j),
\end{aligned}
\end{equation}
where $\odot$ denotes the Hadamard product, and $\vec{w}$ is a trainable weight vector 
to adaptively highlight different dimensions of the item embeddings ${\vec{h}}_i$ and ${\vec{h}}_j$. 
Note that the learned graph structure changes continuously with the update of item embeddings.

To increase the expressive power and stabilize the learning process, 
the similarity metric function can be extended to the multi-head metric~\cite{IDGL, vaswani2017attention}.
Specifically, we use $\phi$ (the number of heads) weight vectors to compute $\phi$ independent similarity matrices (each one representing one perspective) using the above similarity metric function and take their average as the final similarity: 
\begin{equation}\label{eq:similarity_metric_learning}
\begin{aligned}
  M_{ij}^\delta = \text{cos}({\vec{\mathbf{w}}}_\delta \odot {\vec{h}}_i, {\vec{\mathbf{w}}}_\delta \odot {\vec{h}}_j),\quad 
  M_{ij} = \frac{1}{\delta}\sum_{\delta=1}^{\phi}{M_{ij}^\delta},
\end{aligned}
\vspace{-0.1cm}
\end{equation}
where $M_{ij}^\delta$ computes the similarity metric between the two item embeddings $\vec{h}_i$ and $\vec{h}_j$ for the $\delta$-th head, and each head implictly capture different perspective of semantics.

\subsubsection{\textbf{Graph sparsification via $\varepsilon$-sparseness}.}
Typically, the adjacency matrix elements should be non-negative,
but the cosine value $M_{ij}$ calculated from the metric ranges between $[-1, 1]$.
Simply normalizing it does not impose any constraints on the graph sparsity and can yield a fully connected adjacency matrix.
This is computationally expensive and might introduce noise (i.e., unimportant edges), 
and it is not sparse enough that subsequent graph convolutions cannot focus on the most relevant aspects of the graph. 

Therefore, we extract the symmetric sparse non-negative adjacency matrix $A$ from $M$ 
by considering only the node pair with the most vital connection.
To make the hyperparameter of the extraction threshold insensitive and not destroy the graph's sparsity distribution, 
we adopt a \textit{relative ranking strategy of the entire graph}.
Specifically, we mask off (i.e., set to zero) those elements in $M$ that are smaller than a non-negative threshold,
which is obtained by ranking the metric value in $M$.
\begin{equation}\label{eq:epsilon_neigh}
\begin{aligned}
  A_{ij} =
\left\{
        \begin{array}{ll}
             1, & \quad  M_{ij} >   = \mathbf{Rank}_{\varepsilon n^2}(M); \\
              0, & \quad \text{otherwise};
        \end{array}
    \right.
\end{aligned}
\end{equation}
where $\mathbf{Rank}_{\varepsilon n^2}(M)$ returns the value of the $\varepsilon n^2$-th largest value in the metric matrix $M$. 
$n$ is the number of nodes and $\varepsilon$ controls the overall sparsity of the generated graph.

It is different from the \textit{absolute threshold strategy of the entire graph}~\cite{IDGL} 
and the \textit{relative ranking strategy of the node neighborhood}~\cite{graphVQA,chen2019reinforcement}.
The former sets an absolute threshold to remove smaller elements in the adjacency matrix. 
When the hyperparameters are set improperly, as the embedding is continuously updated, 
the metric value distribution will also change,
and it may not be possible to generate a graph or generate a complete graph.
The latter returns the indices of a fixed number of maximum values of each row in the adjacency matrix, 
which will make each node of the generated graph have the same degree. 
Forcing a uniform sparse distribution will make 
the downstream GCN unable to fully utilize the graph's dense or sparse structure information.

\subsection{Interest-fusion Graph Convolutional Layer}\label{ssec::fusion}

As mentioned above, we have learnable interest graphs which separate diverse interests.
The core interests and peripheral interests form large clusters and small clusters respectively, 
and different types of interests form different clusters. 
Furthermore, to gather weak signals to strong ones that can accurately reflect user preferences, we need to aggregate information in the constructed graph.

\subsubsection{\textbf{Interest fusion via graph attentive convolution}}

We propose a \textit{cluster- and query-aware graph attentive convolutional layer}
that can perceive the user's core interest (i.e., the item located in the cluster center) 
and the interest related to query interest (i.e., current target item) during information aggregation. 
The input is a node embedding matrix $\{\vec{h}_1, \vec{h}_2, \dots, \vec{h}_n\}$,
$\vec{h}_i \in \mathbb{R}^d$, 
where $n$ is the number of nodes (i.e., the length of the user interaction sequence), and $d$ is the dimension of embeddings in each node. 
The layer produces a new node embedding matrix
$\{\vec{h}_1', \vec{h}_2', \dots, \vec{h}_n'\}, \vec{h}_i' \in \mathbb{R}^{d'}$, as its output
with potentially different dimension $d'$.

An alignment score $E_{ij}$ is computed to map the importance of target node $v_{i}$ on it's neighbor node $v_{j}$. 
Once obtained, the normalized attention coefficients are used to perform a weighted combination of the embeddings corresponding to them, to serve as the refined output embeddings for every node after applying a residual connection and a nonlinearity function $\sigma$:
\begin{equation}\label{eqnatt}
  \vec{h}'_i =  \sigma\left({\bf{W}_a} \cdot {\mathbf{Aggregate}} \left ({E_{ij} * \vec{h}_j | j\in\mathcal{N}_i} \right) + \vec{h}_i \right). 
\end{equation}

Note that aggregation function can be a function such as \texttt{Mean}, \texttt{Sum}, \texttt{Max}, \texttt{GRU}, etc. We use the simple sum function here and leave other functions for future exploration.
To stabilize the attention mechanism's learning process,
we employ multi-head attention similar to~\cite{vaswani2017attention, GAT}. 
Precisely, $\phi$ independent attention mechanisms execute the above transformation, 
and then their embeddings are concatenated as the following output representation:
\begin{equation}
  \vec{h}'_i = \concat_{\delta=1}^\phi \sigma\left({\bf{W}_a}^\delta \cdot {\mathbf{Aggregate}} \left ({E_{ij}^\delta * \vec{h}_j | j\in\mathcal{N}_i} \right) + \vec{h}_i \right),
\end{equation}
where $\parallel$ represents concatenation operation, $E_{ij}^{\delta}$ are normalized attention coefficients obtained by the $\delta$-th attention head, 
and ${\bf{W}_a}^{\delta}$ is the corresponding linear transformation's weight matrix. 
It is worth noting that the final returned output 
$\vec{h}'$ will correspond to $\phi d'$ dimension embeddings (rather than $d'$) for each node.

\subsubsection{\textbf{Cluster- and query-aware attention}.}
To strengthen important signals and weaken noise signals when integrating interests,
we propose a cluster and query-aware attention mechanism.
We uses the attention coefficients to redistribute weights on edge information in the process of message passing.
The attention mechanism considers the following two aspects.

Firstly, we assume that the target node $v_{i}$'s neighborhood will form a cluster
and regard the target node $v_{i}$ in the graph as a medoid of a cluster $c(v_{i})$.
We define the k-hop neighborhood of the target node $v_{i}$ as the receptive field of the cluster. 
The average value of all nodes' embedding in the cluster $\vec{h}_{i_c}$ represents the cluster's average information.
To identify whether the target node is the center of the cluster,
the target node embedding and its cluster embedding are used to calculate the following attention score,
\begin{equation}
\begin{aligned}
    & \alpha_{i} = \mathbf{Attention}_c({\bf{W}_c} \vec{h}_{i}  \mathbin{\Vert} \vec{h}_{i_c} \mathbin{\Vert}  {\bf{W}_c} \vec{h}_{i} \odot \vec{h}_{i_c}), \\
\end{aligned}
\end{equation}
where ${\bf W_c}$ is a transformation matrix, $\mathbin\Vert$ is the concatenation operator and $\odot$ denotes the Hadamard product.
In our experiments, the attention mechanism $\mathbf{Attention}_c$ is a two-layers feedforward neural network with the \textit{LeakyReLU} as activation function.

Secondly, in order to serve the downstream dynamic pooling method 
and learn the user interest's independent evolution for different target interests,
the correlation between the source node embedding $\vec{h}_{j}$ and the target item embedding $\vec{h}_{t}$ should also be considered. 
If the source node is more correlated with the query item, 
its weight in the aggregation towards the target node will be more significant, and vice versa.
Since only relevant behaviors can play a role in the final prediction, 
we only keep relevant information, and irrelevant information will be discarded during aggregation.
\begin{equation}
\begin{aligned}
    & \beta_{j} = \mathbf{Attention}_q({\bf{W}_q} \vec{h}_{j}  \mathbin\Vert \vec{h}_{t} \mathbin\Vert {\bf{W}_q} \vec{h}_{j} \odot \vec{h}_{t}), \\
\end{aligned}
\end{equation}
where ${\bf{W}_q}$ is a transformation matrix, $\mathbin\Vert$ is the concatenation operator and $\odot$ denotes the Hadamard product.
In our experiments, the attention mechanism $\mathbf{Attention}_q$ is a two-layers feedforward neural network 
applying the LeakyReLU nonlinearity.

We follow the additive attention mechanism~\cite{bahdanau} to consider the factors of cluster and query simultaneously.
We sum the target node's cluster score and the source node's query score as the update weight of the source node $j$ to the target node $i$.
To make coefficients easily comparable across different nodes, 
we employed the softmax function to normalize them across all choices of $j$.
The \emph{attention coefficients} $E_{ij}$ is computed as:
\begin{equation}
\label{eq:s2t-add}
    E_{ij} = \mathrm{softmax}_j(\alpha_{i} + \beta_{j})=\frac{\exp(\alpha_{i} + \beta_{j})}{\sum_{k\in\mathcal{N}_i}{\exp(\alpha_{i} + \beta_{k})}},
\end{equation}
where neighborhood $\mathcal{N}_i$ of node $i$ includes node $i$ itself.
In the context of containing self-loop propagation (when $j$ equals $i$),
$\alpha_{i}$  controls how much information the target node can receive, and $\beta_{j}$ controls how much information the source node can send.

\subsection{Interest-extraction Graph Pooling Layer}

The fusion of implicit interest signals to explicit interest signals is completed by performing information aggregation on the interest graph.
In this section, we use the graph pooling method~\cite{diffpool, sag, asap} to further extract the fused information. 
Similar to the downsampling of feature maps in Pooling in CNN, graph pooling aims to downsize the graph reasonably.
Through the coarsening of the constructed graph structure, 
loose interest is transformed into tight interest and its distribution is maintained.

\subsubsection{\textbf{Interest extraction via graph pooling}}

To obtain the pooled graph, a cluster assignment matrix is necessary~\cite{diffpool, asap}. 
Assuming that a soft cluster assignment matrix $S\in \mathbb{R}^{n \times m}$ exists, it can pool node information into cluster information.
$m$ is a pre-defined model hyperparameter that reflects the degree of pooling, where $m<n$.
Given the node embeddings $\{\vec{h}'_1, \vec{h}'_2,\ldots, \vec{h}'_n\}$ and the node scores $\{\gamma_1, \gamma_2,\ldots, \gamma_n\}$ of the raw graph, 
the cluster embeddings and scores of the coarsened graph can be generated as follows,
\begin{align}
  \{\vec{h}^*_1, \vec{h}^*_2,\ldots, \vec{h}^*_m\} &=S^{T} \{\vec{h}'_1, \vec{h}'_2,\ldots, \vec{h}'_n\}, \\
  \{\gamma^*_1, \gamma^*_2,\ldots, \gamma^*_m\} &=S^{T} \{\gamma_1, \gamma_2,\ldots, \gamma_n\},
\end{align}
where $\gamma_{i}$ obtained by applying softmax on $\beta_{i}$ represents importance score of the $i$-th node.
Each row of assignment matrix $S$ corresponds to one of the $n$ nodes, and each column corresponds to one of the $m$ clusters. 
It provides a soft assignment of each node to the corresponding cluster.
Above equations aggregate node embeddings and scores according to the cluster assignment $S$, 
thereby generating new embedding and score for each of the $m$ clusters.

Next, we discuss how to learn differentiable soft clusters assignment $S$ for nodes.
We use the GNN architecture\cite{diffpool} to generate the assignment matrix. 
The probability matrix of the assignment mapping is obtained through standard message passing and the softmax function, 
based on the adjacency matrix and the node embedding.
\begin{align}
  S_{i:} &= \operatorname{softmax} \left({\bf{W}_p} \cdot {\mathbf{Aggregate}} \left ({A_{ij} * \vec{h}'_j | j\in\mathcal{N}_i} \right) \right),
\end{align}
where the output dimension of weight matrix $\bf{W}_p$ corresponds to the maximum number of clusters $m$.
The softmax function is used to obtain the probability of the $i$-th node being divided into one of $m$ clusters.
It is worth noting that we can obtain the adjacency matrix $A^*$ of the pooled graph by performing $S^{T}AS$, 
ensuring the connectivity between clusters. 
Then, the repetition of the above equations can perform multi-layer pooling to achieve hierarchical compression of interest.

\subsubsection{\textbf{Assignment regularization.}}\label{sec::regularization}
However, it is difficult to train the cluster assignment matrix $S$ using only the gradient signal from the downstream recommendation task. 
The non-convex optimization problem makes it easy to fall into the local optimum in the early training stage\cite{diffpool}.
In addition, the relative position of each node embedding in $\{\vec{h}'_1, \vec{h}'_2,\ldots, \vec{h}'_n\}$ corresponds to the temporal order of the interaction. 
But in the pooled cluster embedding matrix $\{\vec{h}^*_1, \vec{h}^*_2,\ldots, \vec{h}^*_m\}$, the temporal order between the clusters reflecting the user's interest is difficult to be guaranteed.
Therefore, we use three regularization terms to alleviate the above issue.

\begin{itemize}[leftmargin=*]
  \item \textBF{Same mapping regularization.}
To make it easier for two nodes with greater connection strength to be mapped to the same cluster, the first regularization is used as follows,
\begin{align}
  L_{\mathrm{M}} &=\|A, SS^{T}\|_{F},
\end{align}
where $\| \cdot \|_{F}$ denotes the Frobenius norm.
Each element in adjacency matrix $A$ represents the connection strength between two nodes, and each element in $SS^{T}$ represents the probability that two nodes are divided to the same cluster.

  \item \textBF{Single affiliation regularization.}
To clearly define the affiliation of each cluster, we make each row $S_{i:}$ in assignment matrix approach a one-hot vector 
by regularizing the entropy as follows,
\begin{align}
  L_{\mathrm{A}} &=\frac{1}{n} \sum_{i=1}^{n} H\left(S_{i:}\right),
\end{align}
where $H(\cdot)$ is the entropy function that can reduce the uncertainty of the mapping distribution.
The optimal situation is that the $i$-th node is only mapped to one cluster, and the entropy $H(S_{i:})$ is 0 at this time.

  \item \textBF{Relative position regularization.}
The temporal order of the user's interest before and after pooling needs to be maintained for downstream interest evolution modeling.
However, the operation of swapping the index on the pooled cluster embedding matrix $\{\vec{h}^*_1, \vec{h}^*_2,\ldots, \vec{h}^*_m\}$ is not differentiable.
Therefore, we design a position regularization to ensure the temporal order between clusters during pooling as follows,
\begin{align}
  L_{\mathrm{P}} &=\|P_{n}S, P_{m}\|_{2},
\end{align}
where $P_n$ is a position encoding vector $\{1,2,\ldots,n\}$, and $P_m$ is a position encoding vector $\{1,2,\ldots,m\}$. 
Minimizing the L2 norm makes the position of the non-zero elements in $S$ closer to the main diagonal elements.
Intuitively, for the node with the front position in the original sequence, the position index of the cluster to which it is assigned tends to be in the front.
\end{itemize}

\subsubsection{\textbf{Graph readout.}}
At this point, we have obtained a tightly coarsened graph $\mathcal{G}^*$ representing the user's stronger interest signal. 
At the same time,
we perform a weighted readout on raw graph $\mathcal{G}$ to constrain each node's importance,
which aggregates all node embeddings after the forward computation of the propagation layer to generate a graph-level representation ${\vec h}_g$:
\begin{equation}
  \vec h_g=\mathbf{Readout}(\{\gamma_{i} * \vec{h}_i',i\in \mathcal{G}\}),
\end{equation}
where the weight is the score $\gamma_i$ of each node before pooling, and the Readout function can be a function such as \texttt{Mean}, \texttt{Sum}, \texttt{Max}, etc. 
We use the simple sum function here to ensure permutation invariant and leave other functions for future exploration. 
We feed this graph-level representation into the final prediction layer to better extract each cluster's information in the pooling layer.

\subsection{Prediction Layer}\label{i2i-prob-sec}

\subsubsection{\textbf{Interest evolution modeling.}}
Under the joint influence of the external environment and internal cognition, the users' core interests are continually evolving.
The user may become interested in various sports for a time and need books at another time.
However, only using the readout operation mentioned above does not consider the evolution between core interests, which will undoubtedly cause the time order's bias.
To supply the final representation of interest with more relative historical information, it is also necessary to consider the chronological relationship between interests.

Benefiting from the relative position regularization, the pooled cluster embedding matrix maintains the temporal order of the user's interest,
which is equivalent to flattening the pooled graph into a reduced sequence with enhanced interest signals.
Intuitively, we can use any known sequential recommendation method to model the concentrated interest sequence.
For the sake of simplicity and to illustrate the effectiveness of the pooling method, we use a single sequential model to model the evolution of interest:
  \begin{eqnarray}\label{eq:gru1}
    \vec h_s  = \mathbf{AUGRU} (\{\vec{h}^*_1, \vec{h}^*_2,\ldots, \vec{h}^*_m\}).
  \end{eqnarray}
As we know, GRU overcomes the vanishing gradients problem of RNN and is faster than LSTM~\cite{LSTM}.
Furthermore, 
to make better use of the importance weight $\gamma_i^*$ of fused interest in the \textit{interest extraction layer}, 
we adopt GRU with attentional update gate (AUGRU)~\cite{DIEN} to combine attention mechanism and GRU seamlessly.
AUGRU uses attention score $\gamma_i^*$ to scale all dimensions of the update gate, which results that less related interest make fewer effects on the hidden state. 
It avoids the disturbance from interest drifting more effectively and pushes the relative interest to evolve smoothly.  

\subsubsection{\textbf{Prediction.}}

We take the graph-level representation of the \textit{interest extraction layer} and evolution output of the \textit{interest evolution layer} as the user's current interest, 
and concatenate them with the target item embedding.
Given the concatenated dense representation vector, fully connected layers are used to automatically learn the combination of embeddings.
We use two-layer feedforward neural network as the prediction function to estimate the probability of the user interacting with the item at the next moment, and all compared models in the experimental part will share this popular design~\cite{DIN, DIEN, SLIREC},
\begin{equation}
  \hat{y} = \mathbf{Predict}({\vec h}_s \| {\vec h}_g \| {\vec h}_t \| {\vec h}_g \odot {\vec h}_t).
\end{equation}

Following the CTR (click-through rate) prediction in the real-world industry~\cite{DIN, DIEN}, 
we use the negative log-likelihood function as the loss function and share this setting with all compared models.
The optimization process is to minimize the loss function together with a $L2$ regularization term to prevent over-fitting, 
\begin{equation}
  L= - \frac{1}{|\mathcal{O}|} \sum_{\bs{o}\in\mathcal{O}} (y_o\log {\hat{y}}_o + (1-y_o)\log(1-{\hat{y}}_o)) + \lambda \|\Theta\|_2, 
\end{equation}
where $\mathcal{O}$ is the training set and $|\mathcal{O}|$ is the number of training instances.
$\Theta$ denotes the set of trainable parameters and $\lambda $ controls the penalty strength.
The label $y_o = 1$ indicates a positive instance and $y_o = 0$ indicates a negative instance. 
And ${\hat{y}}_o$ stands for the network's output after the softmax layer, representing the predicted probability of the next item being clicked.
Besides, the three regularization terms in Section~\ref{sec::regularization} are added to the final recommendation objective function 
to obtain the better performance and more interpretable cluster assignments.

\section{Experiment}\label{sec::experiments}

In this section, we conduct experiments on two real-world  datasets for sequential recommendation to evaluate our proposed method, with the purpose of answering the following three questions.

\begin{itemize}[leftmargin=*,partopsep=0pt,topsep=0pt]
    \setlength{\itemsep}{0pt}
    \setlength{\parsep}{0pt}
    \setlength{\parskip}{0pt}
\item \textbf{RQ1:} 
  How does the proposed method perform compared with state-of-the-art sequential recommenders?
\item \textbf{RQ2:} 
  Can the proposed method be able to handle sequences with various length effectively and efficiently?
\item \textbf{RQ3:} 
  What is the effect of different components in the method?
\end{itemize}

\subsection{Experimental Settings}

\subsubsection{\textbf{Dataset}}

We evaluate the recommendation performance on a public e-commerce dataset and an industrial short-video dataset.
Table~\ref{tab::dataset} summarizes the basic statistics of the two datasets. 
Average Length represents the average of users' history length, which indicates that the scale of the industry dataset we adopt is much larger than the public dataset.
\begin{itemize}[leftmargin=*]
	\item \textbf{Taobao\footnote{\url{https://tianchi.aliyun.com/dataset/dataDetail?dataId=649}}.} This dataset is widely used for recommendation research \cite{zhu2019joint, pi2019practice}, which is collected from the largest e-commerce platform in China. We use the click data from November 25 to December 3, 2017 and filter out users with less than 10 interactions. We use the first 7 days as training set, the 8th day as validation set, and the last day as test set.
	\item \textbf{Kuaishou\footnote{\url{https://www.kuaishou.com/en}}.} This is an industrial dataset collected from one of the largest short-video platforms in China. Users can upload short-videos and browse other users' short-videos. We downsample the logs from October 22 to October 28, 2020. User behaviors such as click, like, follow (subscribe) and forward are recorded in the dataset. Click data is used to conduct experiments, and the 10-core setting is also adopted to filter out inactive users and videos. Behaviors of the first 6 days are used to train recommendation models. Behaviors during the before 12 pm of the last day is used as validation set, and we keep the instances after 12 pm of the last day to evaluate the final recommendation performance.
\end{itemize}

\begin{table}
	\caption{Statistics of the Datasets.}\label{statistics}
	\vspace{-0.2cm}
	\label{tab::dataset}
	\begin{tabular}{cccccc}
    \toprule
    Dataset & Users & Items & Instances & Average Length \\ 
    \midrule
    Taobao & 36,915 & 64,138 & 1,471,155 & 39.85 \\
    Kuaishou & 60,813 & 292,286 & 14,952,659 & 245.88 \\
    \bottomrule
    \end{tabular}
    \vspace{-0.2cm}
\end{table}

\subsubsection{\textbf{Evaluation Metrics}}
To evaluate the performance of each model, we use two widely adopted accuracy metrics including AUC and GAUC~\cite{DIN}, 
as well as two ranking metrics MRR and NDCG\@. They are defined as follows,
\begin{itemize}[leftmargin=12pt]
    \item \textbf{AUC} signifies the probability that the positive item sample's score is higher than the negative item sample's score, 
      reflecting the classification model's ability to rank samples.
    \item \textbf{GAUC} performs a weighted average of each user's AUC, where the weight is his number of clicks.
    It eliminates the bias between users and evaluates model performance with a finer granularity.
    \item \textbf{MRR} is the mean reciprocal rank, which is the mean value of the inverse of the ranking of the first hit item. 
    \item \textbf{NDCG@K} assigns higher scores to hits at higher positions in the top-K ranking list, which emphasizes that test items should be ranked as higher as possible. In our experiments, we set K to 2, a widely-used setting in existing works.
\end{itemize}

\begin{table*}[ht]
\renewcommand\arraystretch{0.9}
\center
\caption{Performance comparisons
(\textbf{bold} means \textit{p}-value < 0.05, \textbf{bold}* means \textit{p}-value < 0.01, and \textbf{bold}** means \textit{p}-value < 0.001.) }
\label{tab::performance}
{\begin{tabular}{lcccccccccc}
\toprule
\multirow{3}{*}{\textBF{Method}} & &\multicolumn{4}{c}{\textBF{Taobao}} & &\multicolumn{4}{c}{\textBF{Kuaishou}} \\
\cmidrule(lr){3-6}  \cmidrule(lr){8-11} 
&
& \textBF{AUC}   & \textBF{GAUC}
& \textBF{MRR}   & \textBF{NDCG@2}
&
& \textBF{AUC}   & \textBF{GAUC}
& \textBF{MRR}   & \textBF{NDCG@2} \\
\midrule
NCF 
&  & 0.7128 & 0.7221 & 0.1446 & 0.0829      &   & 0.5559 & 0.5531 & 0.7734 & 0.8327       \\
DIN 
&  & 0.7637 & 0.8524 & 0.3091 & 0.2352      &   & 0.6160 & 0.7483 & 0.8863 & 0.9160       \\
LightGCN 
&  & 0.7483 & 0.7513 & 0.1669 & 0.1012      &   & 0.6403 & 0.6407 & 0.8175 & 0.8653       \\
\midrule
Caser 
&  & 0.8312 & 0.8499 & 0.3508 & 0.2890      &   & 0.7795 & 0.8097 & \underline{0.9100} & \underline{0.9336}       \\
GRU4REC 
&  & \underline{0.8635} & \underline{0.8680} & \underline{0.3993} & \underline{0.3422}        &  & \underline{0.8156} & \underline{0.8333} & \underline{0.9174} & \underline{0.9391}   \\
DIEN 
&  & 0.8477 & \underline{0.8745} & \underline{0.4011} & \underline{0.3404}                    &  & 0.7037 & 0.7800 & 0.9030 & 0.9284   \\
SLi-Rec
&  & \underline{0.8664} & 0.8669 & 0.3617 & 0.2971                                            &   & \underline{0.7978} & \underline{0.8128} & 0.9075 & 0.9318  \\
\midrule
\textBF{SURGE}
&  & $\textBF{0.8906}^{**}$ & $\textBF{0.8888}^{}$ & $\textBF{0.4228}^{*}$ & $\textBF{0.3625}^{**}$      & & $\textBF{0.8525}^{**}$ & $\textBF{0.8610}^{**}$ & $\textBF{0.9316}^{**}$ & $\textBF{0.9495}^*$  \\
\bottomrule            
\end{tabular}
}
\vspace{-0.2cm}
\end{table*}

\subsubsection{\textbf{Baselines}}
To demonstrate the effectiveness of our SURGE model, we compare it with competitive sequential recommenders. 
The baselines are classified into two categories: non-sequential model that only captures user's static interest, and sequential models that consider dynamic interest patterns.

\noindent\textbf{Non-sequential Models:}
\begin{itemize}[leftmargin=*,partopsep=0pt,topsep=0pt]
\setlength{\itemsep}{0pt}
\setlength{\parsep}{0pt}
\setlength{\parskip}{0pt}
  \item \textbf{NCF} \cite{NCF}: This method combines matrix factorization and multi-layer perceptrons to predict user interactions, and it is the state-of-the-art general recommender.
  \item \textbf{DIN}~\cite{DIN}: This method uses attention mechanism with the target item as the query vector. Representation of the user is obtained by aggregating the history interaction with the attention weights.
  \item \textbf{LightGCN} \cite{LightGCN}: This is the state-of-the-art model which uses graph neural network to extract higher-order connectivity for the recommendation.
\end{itemize}
\noindent\textbf{Sequential Models:}
\begin{itemize}[leftmargin=*,partopsep=0pt,topsep=0pt]
\setlength{\itemsep}{0pt}
\setlength{\parsep}{0pt}
\setlength{\parskip}{0pt}
	\item \textbf{Caser}~\cite{Caser}: This method embeds a set of recent item sequences in time and latent space into an image feature and uses convolution filters to learn the sequence patterns.
  \item \textbf{GRU4REC}~\cite{GRU4REC}: This method uses GRU to model user session sequences and encode user interest into a final state.
	\item \textbf{DIEN}~\cite{DIEN}: This method uses a two-layer GRU composed of interest extraction layer and interest evolution layer to model the user's behavior sequence.  
	\item \textbf{SLi-Rec}~\cite{SLIREC}: This is the state-of-the-art method that jointly models long and short-term interests based on an attention framework and an improved time-aware LSTM.
\end{itemize}

It is worth noting that \textit{session recommendation} is another recommendation task similar to sequential recommendation, which aims to predict the next item based on \textbf{only} the user's current session data without utilizing the long-term preference profile.
Recently, graph-based models~\cite{SRGNN, FGNN, GCSAN, GCEGNN} achieve successes on this task. The complex transitions between repeated behaviors in each session are modeled through the small item graphs for each user.
However, users rarely produce repetitive behaviors over a long time, making relevant work impossible to apply to the task of sequential recommendation.

\subsubsection{\textbf{Hyper-parameter Settings.}}\label{param}

We implement all the models with the Microsoft Recommenders framework\footnote{\url{https://github.com/microsoft/recommenders}}
based on TensorFlow\footnote{\url{https://www.tensorflow.org}}. 
We use Adam~\cite{Adam} for optimization with the initial learning rate as 0.001.
The batch size is set as 500 and embedding size is fixed to 40 for all models.
Xavier initialization~\cite{xavier} is used here to initialize the parameters. 
All methods use a two-layer feedforward neural network with hidden sizes of [100, 64] for interaction estimation.
The maximum length for user interaction sequences is 50 for the Taobao dataset and 250 for the Kuaishou dataset.
We apply careful grid-search to find the best hyper-parameters.
All regularization coefficients are searched in $[1e^{-7}, 1e^{-5}, 1e^{-3}]$. 
The pooling length of the user interaction sequence is searched in $[10, 20, 30, 40, 50]$ for Taobao dataset 
and $[50, 100, 150, 200, 250]$ for Kuaishou dataset.

\subsection{Overall Performance (RQ1)}
Table~\ref{tab::performance} illustrates the results on the two datasets.
From the results, we have the following observations:
\begin{itemize}[leftmargin=*,partopsep=0pt,topsep=0pt]
\setlength{\itemsep}{0pt}
\setlength{\parsep}{0pt}
\setlength{\parskip}{0pt}
  \item \textbf{Our proposed method consistently achieves the best performance.}
We can observe that our model SURGE significantly outperforms all baselines in terms of both classification and ranking metrics.
Specifically, our model improves AUC by around 0.03 (\textit{p}-value < 0.001) on Taobao dataset and 0.04 (\textit{p}-value < 0.001) on Kuaishou dataset.
The improvement is more obvious on the Kuaishou dataset with longer interaction history, which verifies that our method can handle long sequences more effectively and significantly reduces the difficulty of modeling user interests.
  \item \textbf{Sequential models are effective but have a short-term bottleneck.} 
Compared with NCF, DIN and LightGCN, the better performance of Caser, DIEN and GRU4Rec verifies the necessity of capturing sequential patterns for modeling user interests.
On the Taobao dataset, RNN-based models (GRU4Rec and DIEN) with more powerful ability to capture sequential patterns outperformed the CNN-based model (Caser).
The max pooling scheme in CNN that is commonly used in computer vision omits important position and recurrent signals when modeling long-range sequence data.
But on the Kuaishou dataset, since RNNs tend to forget long-term interest when processing longer sequences, the performance of DIEN and GRU4Rec are in par with Caser in most metrics. This result indicates that even powerful recurrent neural networks have a short-term memory bottleneck.
In addition, since long sequences tend to contain more noise, DIEN's performance on the two datasets is unstable compared to GRU4REC\@. This shows that the even though two-layer GRU structure is often more effective, the performance is more likely to be impacted by noise on datasets with long sequences,
therefore justifying our motivation to summarize the sequences with metric learning.
  \item \textbf{Joint modeling long and short-term interests does not always add up to better performance.} 
SLi-Rec, which joint models long and short-term interests, is the best baseline on Taobao in terms of the AUC metric, but exhibits poor performance according to ranking metrics.
In addition, on Kuaishou with longer interaction sequences, SLi-Rec's performance is worse than GRU4Rec for all metrics, even though GRU4REC does not explicitly model long and short-term interests.
This indicates that although SLi-Rec utilizes two separate components to model users' long and short-term interests, it still fails to effectively integrate them into a single model, in particular for long sequences.
Moreover, SLi-Rec leverages timestamp information to improve modeling long and short-term interests. However, our method shows better performance by compressing information with metric learning, without the need to explicitly model timestamp.
\end{itemize}

\subsection{Study on Sequence Length and Efficiency Comparison (RQ2)}

\subsubsection{\textbf{Study on Sequence Length.}}\label{param}
In real-world applications, a user may have very long interaction sequences.
Long historical sequences often have more patterns that can reflect user interests, but the accompanying increased noise signals will mislead the modeling of real interests.
Thus, whether to effectively model the user's long-term history is a significant issue for sequential recommendation.
We study how SURGE improves the recommendation for those users with long behavior records.
Specifically, we divide all users of the two datasets into five groups based on the length of the interaction history.
For each group, we compare the performance of our method with the baseline methods and present the GAUC metric of the two datasets, as shown in Figure~\ref{fig:length}.

From the results, we can observe that all models are challenging to capture users' real interest when the sequence length is short due to data sparsity. As the length of the sequence increases and the difficulty of modeling decreases, most models' performance improves and reaches a peak. 
But as the length continues to increase, most models' performance will decline with the introduction of a large number of noise signals.
Among them, DIN and DIEN declined most significantly.
It is difficult for DIN to focus on the most critical parts in a long sequence. 
The item with the greatest attention may occur in the early part of the sequence, and it may be very different from the user's current interest.
When the two-layer GRU structure in DIEN models user interests, the next GRU input depends on the previous GRU output, making it easier to be disturbed by the noise in long sequences.
Due to the short-term bottleneck of a single GRU, GRU4REC will only focus on the recent history and ignore the sequence length, and its performance in each length group is relatively stable.
Although SLi-Rec jointly considers users' long-term and short-term interests, it still models for noise-filled sequences, so it is inevitable to suffer performance degradation on long sequences.

However, the performance gap between SURGE and other methods becomes larger as the sequence length increases.
Furthermore, even in the user group with the longest historical sequence, SURGE still keeps the excellent performance of 0.8919 and 0.8502 on Taobao and Kuaishou datasets, respectively.
Since the SURGE model merges implicit signals into explicit signals and filters out noise, it can achieve good performance for users with a long history.
In summary, we conclude that the SURGE model we proposed can more effectively model users' long-term historical sequence.

\begin{figure}[t]
  \centering
  \hspace{-0.3cm}
  \subfigure{\label{fig:group-taobao}
    \includegraphics[width=0.24\textwidth]{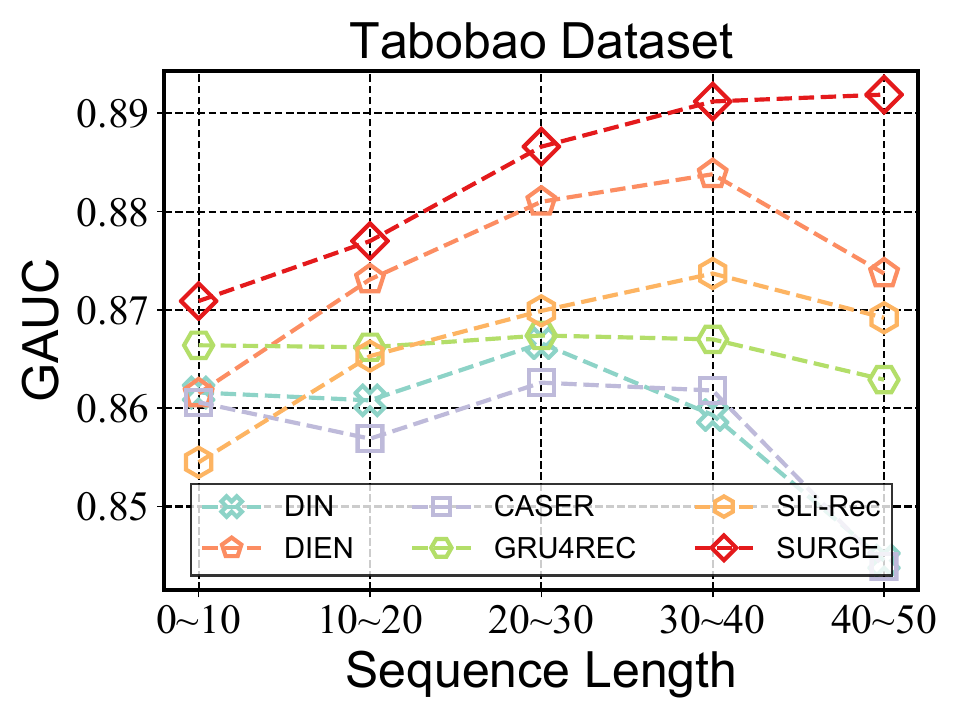}}
  \hspace{-0.2cm}
    \subfigure{\label{fig:group-kuaishou}
    \includegraphics[width=0.24\textwidth]{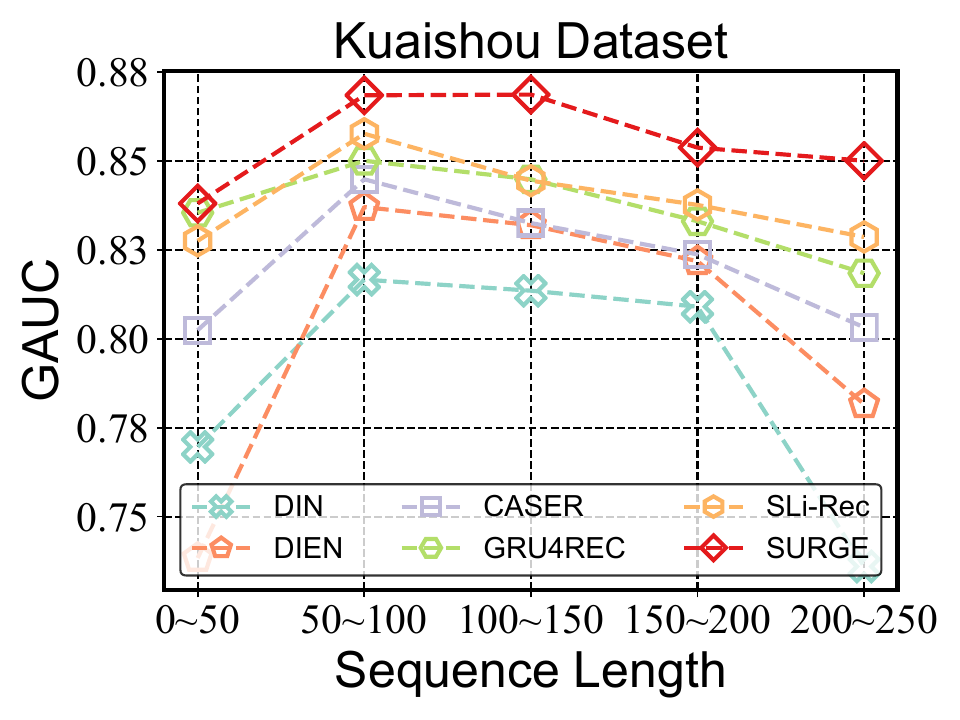}}
  \caption{Performance breakdown by sequence lengths on the two datasets. Best viewed in color.}
  \vspace{-0.4cm}
  \label{fig:group}
\end{figure}

\subsubsection{\textbf{Efficiency Comparison.}}\label{param}

For sequential recommendation systems, it is challenging to efficiently model user behavior history.
The differences and diversity of items in the user's historical sequence lead to slow model convergence.
Besides, long historical sequences often correspond to more complex calculations and more time-consuming training.
To study whether SURGE can alleviate the issue, we visualize the training process of SURGE and baseline models and compare the convergence speed and training time of each model.
Specifically, we plot the performance changes of the proposed method and the baseline methods on the validation set during the training process and reported the GAUC metric.
We use early stop to detect whether the training is over, that is, if the GAUC on the validation set does not increase within five epochs, the training process will stop.
For the two datasets' performance change curves, we use smoothing rates of 0.2 and 0.6 to smooth them to see the trend better.

The training process on the two datasets is shown in Figure 3.
From the results, we can observe that DIN fails to focus on currently activated interests on long sequences, so it continually fluctuates on the kuaishou dataset and it is difficult to converge.
Since GRU4REC is more likely to forget long-term interests, only the item embeddings at the end of the sequence will be updated in each training instance. Therefore, its training curve is steady and slow, and the continuous slight increase makes it hard to stop early.
Because SLIREC specifically considers the long-term interests of users, it converges quickly on the kuaishou dataset, but it is the slowest method on the Taobao dataset with a shorter sequence.

Table~\ref{tab::time} shows each model's training time on the two datasets. 
We can observe that, except for the non-sequential model (DIN) on the kuaishou dataset, our method's efficiency improvement compared with all baselines is more than 20\%.
The reason is that SURGE performs a pooling operation on the sequence before feeding the embedding sequence into the recurrent neural network, which greatly reduces the number of recurrent steps.
Besides, since most of the noise is filtered, the compressed sequence only contains the core interest, which will undoubtedly help speed up the model's convergence.
Therefore, we concluded that the SURGE model can more efficiently model users' long-term historical sequence.

\begin{figure}[t]
  \centering
  \hspace{-0.3cm}
  \subfigure{\label{fig:length-taobao}
    \includegraphics[width=0.24\textwidth]{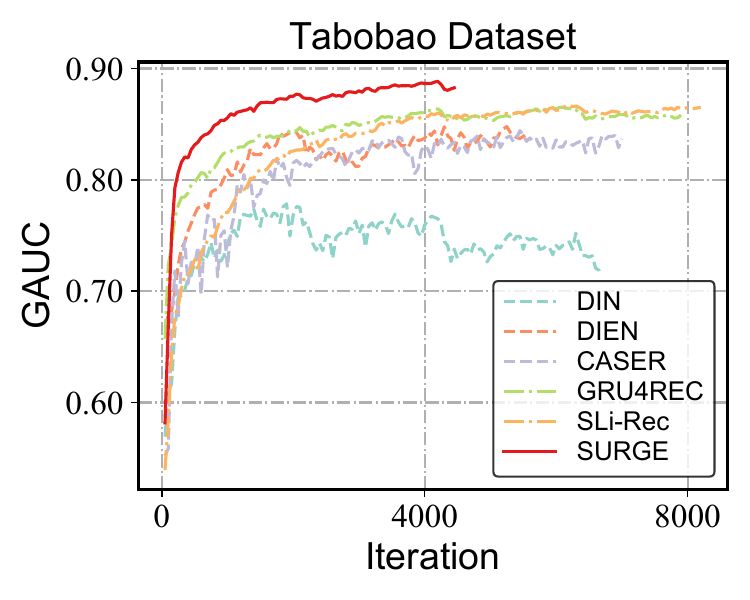}}
   \hspace{-0.3cm}
    \subfigure{\label{fig:length-kuaishou}
    \includegraphics[width=0.24\textwidth]{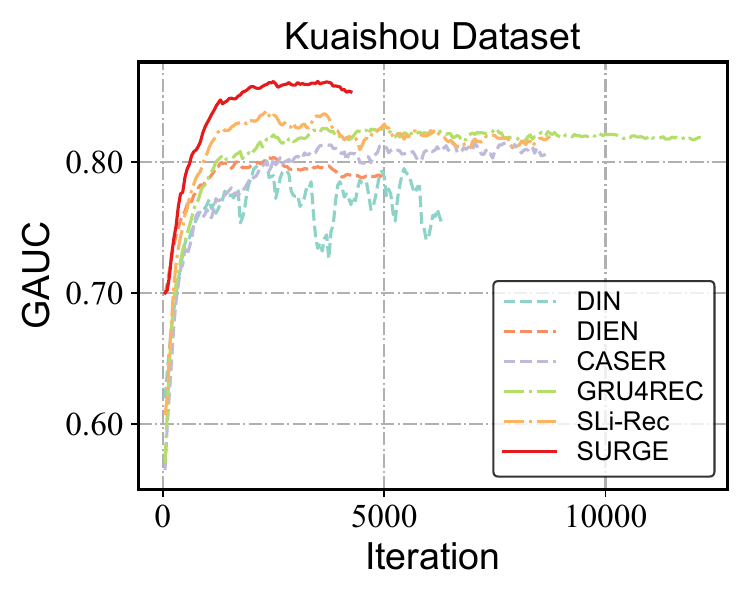}}
  \vspace{-0.4cm}
  \caption{Test performance of the baselines by iterations on two datasets. Best viewed in color.}
  \vspace{-0.1cm}
\label{fig:length}

\end{figure}

\subsection{Ablation and Hyper-parameter Study (RQ3)}

\subsubsection{\textbf{Effectiveness of interest fusion.}}\label{param}

We propose to perform message passing on the interest graph based on similarity to merge weak signals into strong signals. We now investigate whether this fusion design that strengthens core interests and activates target interests is necessary. 
To be specific, we compare the no propagation, cluster-aware propagation, query-aware propagation, cluster- and query-aware propagation.

The results on the two datasets in Table~\ref{tab::ablation} show the effectiveness of fusing weak signals into strong signals through graph convolution.
The enhancement of core interest and the activation of target interest respectively bring further performance improvements.

\begin{table}[t]
\caption{Total training time until convergence of baselines on two real-world datasets, where m indicates minutes.}\label{tab::time}
\vspace{-0.2cm}
\center
\footnotesize
\begin{tabular}{ccccccc}
\toprule
\textBF{Dataset}  
& {DIN}  & {Caser}  & {GRU4REC}  & {DIEN}  & {SLi-Rec}  & {SURGE}  \\ 
\midrule
\textBF{Taobao}  
& 22.65m & 23.66m & 26.78m & 18.74m  & 27.82m & 14.96m        \\ 
\textBF{Kuaishou}   
& 20.59m & 120.26m & 73.35m & 28.47m & 28.84m & 22.86m        \\ 
\bottomrule  
\end{tabular}
\vspace{-0.3cm}
\end{table}

\subsubsection{\textbf{Effectiveness of interest extraction.}}\label{param}

To evaluate the impact of interest extraction through pooling strategy on interest modeling.
We compared no graph pooling, graph pooling without assignment regularization, graph pooling without weighted readout, and complete graph pooling.

The results are shown in Table~\ref{tab::ablation}.
We can observe that interest extraction can help filter irrelevant noise so that the model focuses on the most critical part of modeling.
Especially when the assignment regularization and the graph readout are injected into the model, the user's interest can be better compressed to improve the recommendation performance.

\vspace{-0.1cm}
\subsubsection{\textbf{Design choices for interest evolution.}}\label{param}

\begin{figure}[t]
  \centering
  \hspace{-0.3cm}
  \subfigure{\label{fig:group-taobao}
    \includegraphics[width=0.24\textwidth]{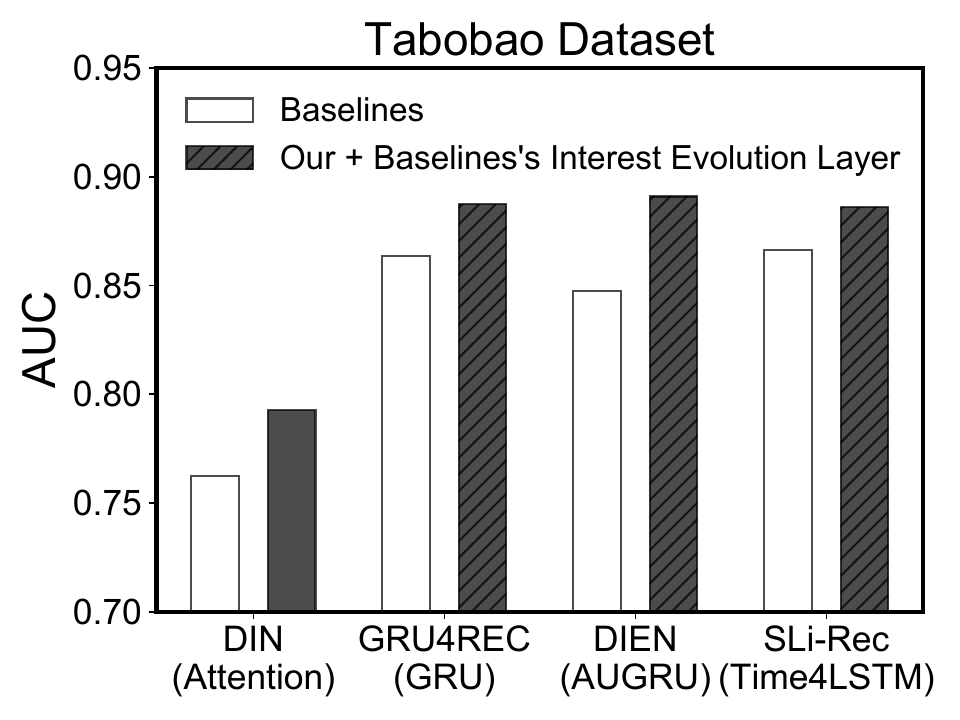}}
   \hspace{-0.2cm}
    \subfigure{\label{fig:group-kuaishou}
    \includegraphics[width=0.24\textwidth]{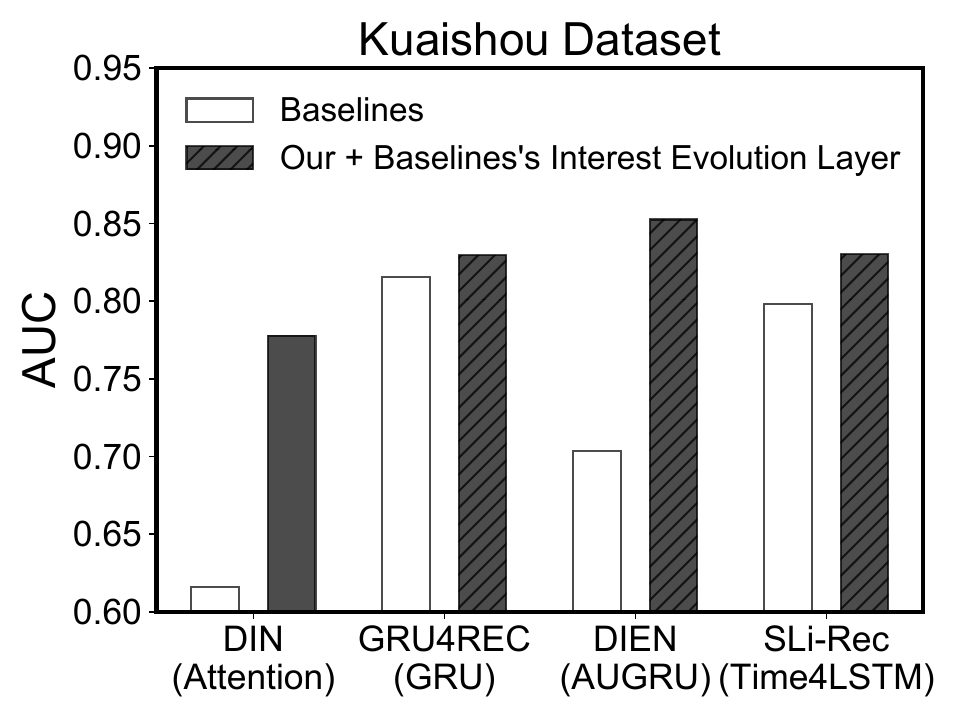}}
  \vspace{-0.4cm}
  \caption{Performance comparison of the proposed method using different interest evolution layers.}
  \vspace{-0.3cm}
  \label{fig:ablation}
\end{figure}

Our framework is agnostic to the selection of the prediction layer after the pooling.
We can use any known sequential recommendation method to model the concentrated interests.
We compared the effects of using different prediction layers on the compressed sequence, including Attention (DIN), GRU (GRU4Rec), AUGRU (DIEN) and TIME4LSTM (SLi-Rec), and the results are shown in Figure~\ref{fig:ablation}.

The first observation is that the performance of sequential models other than DIN are less different, and AUGRU, which can utilize the cluster score in the interest extraction layer, is slightly better.
The second observation is that modeling on the compressed sequence can bring benefits to all existing methods.
It shows that our pooling strategy will significantly reduce the difficulty of modeling user interests and obtain better performance.

In conclusion, we conduct extensive experiments on two real-world datasets, which verifies that our proposed SURGE model outperform existing recommendation methods. Further studies demonstrate our model can effectively and efficiently alleviate the problem that long sequences are difficult to model.

\begin{table}[t]
\small
\centering
\caption{Ablation study of the key designs}
\label{tab::ablation}
\setlength{\tabcolsep}{0.9mm}{\begin{tabular}{llcccccc}
\toprule
\multicolumn{2}{c}{\multirow{3}{*}{\textBF{Model}}}
& &\multicolumn{2}{c}{\textBF{Taobao}} & &\multicolumn{2}{c}{\textBF{Kuaishou}} \\
\cmidrule(lr){4-5}  \cmidrule(lr){6-8} 
&
&
& \textBF{AUC}   & \textBF{MRR}
&
& \textBF{AUC}   & \textBF{MRR}\\
\midrule
\multirow{4}{*}{\makecell[l]{Interest \\ Fusion}
} 
& w/o Fusion
&  & 0.8307 & 0.0317 &  & 0.7149 & 0.9076 \\
&  w/o Query-aware
&  & 0.8720 & 0.3929 &  & 0.7641 & 0.9069 \\
&  w/o Cluster-aware
&  & 0.8764 & 0.3973 &  & 0.8213 & 0.9186 \\
& w/ Fusion
&  & 0.8906 & 0.4228 &  & 0.8525 & 0.9316 \\
\midrule
\multirow{4}{*}{\makecell[l]{Interest \\ Extraction}
} 
& w/o Extraction 
&  & 0.8513 & 0.3605 &  & 0.8240 & 0.9182 \\
& w/o Readout 
&  & 0.8578 & 0.3720 &  & 0.8422 & 0.9257 \\
& w/o Regularization
&  & 0.8815 & 0.3430 &  & 0.8487 & 0.9291 \\
& w/ Extraction 
&  & 0.8906 & 0.4228 &  & 0.8525 & 0.9316 \\
\bottomrule            
\end{tabular}
}
\vspace{-0.3cm}
\end{table}

\section{Related Work}\label{sec::relatedwork}

\para{Sequential Recommendation.}
Sequential recommendation is defined as leveraging historical sequential behaviors for predicting next behavior.
The earliest work, FPMC~\cite{rendle2010factorizing}, used Markov chain to model the transition in behavior sequences.
For stronger ability to learn the complex bahaviors, deep learning-based methods~\cite{DIN, GRU4REC, Caser, DIEN, kang2018self} were proposed, including recurrent neural network-based ones~\cite{DIEN, GRU4REC} and attention network-based ones~\cite{DIN, kang2018self}. These works pay more attention to users' recent behaviors and largely ignore the users' long-term behaviors.
Considering this point, there are some recent works~\cite{zhao2018plastic, SLIREC} combining sequential recommendation model and a normal recommendation model, such as matrix factorization~\cite{koren2009matrix},
to capture long and short-term interest.

However, roughly dividing user interest into long-term part and short-term part is not reasonable.
Compared with these works, in our work, we approach the problem of sequential recommendation from a new perspective: we argue the sequential behaviors reflect weak preference signals, and some part of user preferences may be deactivated at a given time point.

\para{Graph Neural Networks for Recommendation.}
In recommendation scenarios, the input data can be represented in a graph structure.
Recently, with strong ability of learning from graph-structure data, graph neural networks~\cite{GCN, GAT} have become popular means for recommender systems.
PinSage~\cite{Pinsage} applied GCN to pin-board graphs, which was the first work for applying GCN into industrial recommender systems. 
The standard GCN~\cite{GCN} was adopted to factorizing user-item rating matrices into user and item embedding matrices for recommendation~\cite{GCMC} and Wang~\textit{et al.}~\cite{NGCF} further proposed the general solution for implicit recommendation task.
GCN-based methods have achieved the state-of-the-art performance in other recommendation problems, such as social recommendation~\cite{wu2019neural,fan2019graph,zhang2021group}, knowledge graph-based recommendation~\cite{wang2019kgat,wang2019knowledge}, multi-behavior recommendation~\cite{jin2020multi}, 
bundle recommendation~\cite{chang2020bundle}, etc. There are some works~\cite{SRGNN} utilizing graph neural networks for session-based recommendation, a problem similar with sequential recommendation. In session-based recommendation, one of most important goals is to capture the seasonal repetitive behaviors, making it has big difference with sequential recommendation, which is a more general and important problem in the research area.

Differ from aforementioned works, we take advantage of graph convolutional propagation to fuse the weak preference signals to strong ones
and propose graph pooling to extract the dynamically-activated core preference in the long behavior sequences.

\section{Conclusions and Future Work}\label{sec::conclusion}

In this work, we studies the task of sequential recommender systems. We propose a graph-based solution that re-constructs loose item sequences into tight item-item interest graphs.
The model utilizes graph neural network's powerful ability to dynamically fuse and extract users' activated core interests from noisy user behavior sequences.
Extensive experiments on both public and proprietary industrial datasets demonstrate the effectiveness of our proposal. 
Further studies on sequence length confirm that our method can model long behavioral sequences effectively and efficiently.

As for future work, we plan to conduct A/B tests on the online system to further evaluate our proposed solution's recommendation performance.
We also plan to consider using multiple types of behaviors, such as clicks and favorites, to explore fine-grained multiple interactions from noisy historical sequences.

\begin{acks}
This work was supported in part by The National Key Research and Development Program of China under grant 2020AAA0106000, the National Natural Science Foundation of China under U1936217,  61971267, 61972223, 61941117, 61861136003, Beijing Natural Science Foundation under L182038, Beijing National Research Center for Information Science and Technology under 20031887521, and research fund of Tsinghua University - Tencent Joint Laboratory for Internet Innovation Technology.
\end{acks}

\bibliographystyle{ACM-Reference-Format}
\setstretch{1.00}
\balance
\bibliography{bib_full_name}
\end{document}